# Experimental and Theoretical Studies of the Gas-Phase Reactions of O($^1$D) with H$_2$O and D$_2$O at Low Temperature

Kevin M. Hickson,[a,b] Somnath Bhowmick,[c,d] Yury V. Suleimanov,[c] João Brandão,[e] Daniela V. Coelho[e]

[a]Université de Bordeaux, Institut des Sciences Moléculaires, F-33400 Talence, France

[b]CNRS, Institut des Sciences Moléculaires, F-33400 Talence, France

[c]Computation-based Science and Technology Research Center, The Cyprus Institute, 20 Konstantinou Kavafi Street, Nicosia 2121, Cyprus

[d]Climate & Atmosphere Research Centre, The Cyprus Institute, 20 Konstantinou Kavafi Street, Nicosia 2121, Cyprus

[e]Departamento de Química e Farmácia – FCT, Universidade do Algarve, Campus de Gambelas, 8005-139 Faro, Portugal

**Abstract**

Here we report the results of an experimental and theoretical study of the gas-phase reactions between O($^1$D) and H$_2$O and O($^1$D) and D$_2$O at room temperature and below. On the experimental side, the kinetics of these reactions have been investigated over the 50-127 K range using a continuous flow Laval nozzle apparatus, coupled with pulsed laser photolysis and pulsed laser induced fluorescence for the production and detection of O($^1$D) atoms respectively. Experiments were also performed at 296 K in the absence of a Laval nozzle. On the theoretical side, the existing full-dimensional ground X $^1$A potential energy surface for the H$_2$O$_2$ system involved in this process has been reinvestigated and enhanced to provide a better description of the barrierless H-atom abstraction pathway. Based on this enhanced potential energy surface, quasiclassical trajectory calculations and ring polymer molecular dynamics simulations have been performed to obtain low temperature rate constants. The measured and calculated rate constants display similar behaviour above 100 K, showing little or no variation as a function of temperature. Below 100 K, the experimental rate constants increase dramatically, in contrast to the essentially temperature independent theoretical values. The possible origins of the divergence between experiment and theory at low temperatures are discussed.

# 1 Introduction

The reactions of atomic oxygen in its first electronically excited state, O($^1$D), play important roles in the chemistry of several planetary atmospheres, including Earth and Mars, where oxygen bearing molecules are among the major atmospheric constituents. In these environments, the absorption of solar radiation at ultraviolet (UV) and vacuum ultraviolet (VUV) wavelengths generates a range of reactive radicals. In the Earth's stratosphere, O($^1$D) atoms are generated by O$_3$ photolysis in the UV range, where O$_3$ itself is formed by the termolecular recombination of ground electronic state oxygen atoms O($^3$P) (produced by O$_2$ photolysis) with O$_2$. The subsequent reactions of O($^1$D) atoms with hydrogen bearing molecules such as H$_2$, CH$_4$ and H$_2$O[1] are considered to be among the most important stratospheric sources of the hydroxyl radical (OH) and compete with quenching losses mostly through collisions with N$_2$ and O$_2$.[2] In the Martian atmosphere, O($^1$D) atoms are thought to be formed essentially by the photodissociation of CO$_2$, the major atmospheric component. At VUV wavelengths shorter than 170 nm, CO$_2$ photolysis produces O($^1$D) with high quantum yields.[3] While these atoms are mostly removed by quenching to the ground state through collisions with CO$_2$,[4] a small fraction of these atoms react with H$_2$ and H$_2$O, present at trace levels generating OH radicals.[5] The subsequent reaction of OH radicals with CO acts to regenerate CO$_2$, helping to maintain the stability of the Martian atmosphere.

Temperature dependent kinetic studies of the O($^1$D) + H$_2$ reaction[6-8] and the O($^1$D) + CH$_4$ reaction[9-12] have demonstrated that these processes remain rapid down to low temperature, while OH has been shown to be a major product of the O($^1$D) + CH$_4$ reaction.[9, 12] The reaction of O($^1$D) with H$_2$O has several exothermically accessible channels

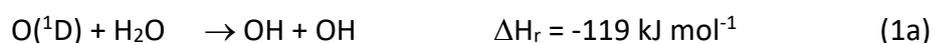
O($^1$D) + H$_2$O  → OH + OH         ΔH$_r$ = -119 kJ mol$^{-1}$         (1a)

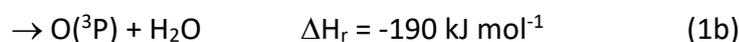
→ O($^3$P) + H$_2$O        ΔH$_r$ = -190 kJ mol$^{-1}$         (1b)

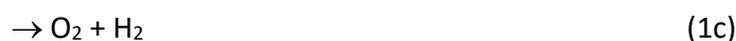
→ O$_2$ + H$_2$                                            (1c)

where O$_2$ in channel (1c) can be formed in several different electronic states leading to values for ΔH$_r$ between -40 and - 197 kJ mol$^{-1}$.[6] This process has been studied over a wide temperature range (217-453 K) by various groups.[6, 13, 14] These studies have shown that the rate constants are large and essentially constant with measured rate constants between 2.0 and 2.5 × 10$^{-10}$ cm$^3$ s$^{-1}$, while the dominant (almost exclusive) product channel is (1a) leading to OH + OH.[14-17] On the theoretical side, based on ab initio calculations of the ground $^1$A potential energy surface (PES) using the Moller-Plesset perturbation theory, Sayos et al.[18] derived rate constants for this reaction at 300 K through quasiclassical trajectory (QCT) calculations obtaining a rate constant value (corrected for the electronic degeneracy factor) of 1.32 × $10^{-10}$ $cm^3$ $s^{-1}$; somewhat lower than the measured ones at room temperature. Despite the relatively large body of previous work on the $O(^1D)$ + $H_2O$ reaction, there are no measurements below 217 K, presumably because most previous work has focused on its importance for Earth's stratospheric ozone chemistry, where temperatures are typically greater than 200 K. Another major obstacle to the study of gas-phase reactions involving $H_2O$ at even lower temperature arises from its low saturated vapour pressure, placing severe constraints on the useful range of $H_2O$ concentrations. One technique that has already been applied to investigate the gas-phase reactivity of $H_2O$ below 200 K is the CRESU method (where CRESU stands for cinétique de réaction en écoulement supersonique uniforme or reaction kinetics in a uniform supersonic flow) as demonstrated by earlier kinetic measurements of the reactions of CH,[19] $C(^3P)$[20] and $C(^1D)$[21] with $H_2O$ down to temperatures as low as 50 K.

In the present work, we report an experimental and theoretical study of the $O(^1D)$ + $H_2O$ and $O(^1D)$ + $D_2O$ reactions over the 50-296 K temperature range to extend the measured data for these reactions down to 50 K. On the experimental side, a CRESU reactor was employed during this investigation coupled with pulsed laser photolysis and pulsed laser induced fluorescence to generate and detect $O(^1D)$ atoms, respectively. On the theoretical side, quantum chemical calculations were performed to better describe the ground singlet ($^1A$) PES of the $H_2O_2$ system involved in the present reaction. This PES was then employed in quasiclassical trajectory (QCT) calculations and path integral based ring polymer molecular dynamics (RPMD) simulations to deduce correlation with the experimental results. The experimental and theoretical methods are described in sections 2 and 3, respectively. Section 4 presents the experimental and theoretical results, which are discussed in section 5. Our conclusions are given in section 6.

**2 Experimental Methods**

The measurements described here were performed using a continuous supersonic flow or CRESU type reactor whose major features can be found in earlier publications.[22, 23] A schematic representation of the apparatus can be found in Hickson et al.[24] The central element of this system is the Laval nozzle, allowing supersonic flows with uniform density, velocity and low temperature profiles to be generated through the isentropic expansion of a carrier gas into a vacuum chamber. Subsequent modifications to the apparatus (notably with respect to the efficient generation of tunable light in the VUV wavelength range in addition to sensitive detection of VUV emission) have allowed us to investigate the kinetics of $O(^1D)$ reactions[2, 4, 8, 12, 25-28] as well as the kinetics of other atomic radical reactions.[4, 20, 21, 29-41] For the present experiments, argon was used exclusively as the carrier gas due to the relatively slow quenching of $O(^1D)$ atoms by Ar,[2, 42] while $N_2$ based nozzles could not be used due to the rapid quenching of $O(^1D)$ atoms by $N_2$.[2, 13] Three different Ar based nozzles were employed here, providing supersonic flows with temperatures of 127 K, 75 K and 50 K. The flow characteristics for these Laval nozzles are reported in Table 1 of Nuñez-Reyes and Hickson.[4] Experiments were also performed without a nozzle, allowing us to measure rate constants at room temperature (296 K).

For most of the experiments described here, water was introduced into the gas flow upstream of the Laval nozzle using a controlled evaporation mixing (CEM) system. A 1 litre stainless steel reservoir maintained at a positive pressure of 2 bar was connected to a liquid flow meter allowing flows of between 0.1 - 5.0 g hr$^{-1}$ of liquid water to be passed into an evaporation device heated to 373 K. A small fraction of the Ar carrier gas flow was diverted into the evaporation system through a mass flow controller, allowing water vapour to be carried into the reactor. To determine the gas-phase water concentration, the output of the CEM was flowed through a **10-cm** absorption cell which was coupled a mercury lamp and a solar blind channel photomultiplier tube (CPM) operating in photon counting mode. To ensure that only radiation from the 185 nm mercury line was detected, an interference filter with a 10 nm FWHM transmission around 185 nm was placed in front of the CPM. The UV radiation transmitted by the cell was measured before, during and after each set of experiments with a given flow of $H_2O$ vapour allowing the attenuated and non-attenuated intensities to be recorded. This procedure also allowed us to estimate the signal drift as a function of experiment time due to changing lamp conditions. The pressure within the cell could be set by adjusting a needle valve placed at the exit of the cell. This was generally maintained at a value around 650 Torr, ensuring an adequate level of $H_2O$ absorption. The room temperature absorption cross-section of water vapour was taken to be $6.78 \times 10^{-20}$ cm$^2$ as recommended by Sander et al.,[15] allowing its concentration to be calculated by the Beer-Lambert law. $H_2O$

condensation downstream of the absorption cell was avoided by connecting the cell exit port to the reactor using a heated hose maintained at 353 K. As the water vapour was diluted by at least a factor of five on entering the nozzle reservoir through mixing with the main carrier gas flow, we assume that no supplementary condensation losses occurred upstream of the Laval nozzle. $H_2O$ concentrations as high as $2.9 \times 10^{14}$ cm$^{-3}$ were used in the cold flow.

$D_2O$ could not be introduced into the reactor using the CEM system as a result of the limited quantity available. Instead, a small fraction of the Ar carrier gas was flowed into a bubbler containing $D_2O$ at room temperature. At the exit of the bubbler, the Ar containing $D_2O$ vapour was passed into a cold trap, held below room temperature at a known pressure. The output of the cold trap was flowed to the Laval nozzle reservoir through the same heated hose described above. Although it was not possible to determine the gas-phase $D_2O$ concentration spectroscopically, as its absorption cross section is an order of magnitude weaker than the $H_2O$ one at 185 nm, the procedure described above ensured that the gas-phase $D_2O$ was maintained at its saturated vapour pressure value[43] at the given temperature below 296 K, allowing the gas-phase [$D_2O$] concentration in the supersonic flow to be calculated precisely. To check for potential errors arising from vapour saturation issues, several experiments with $D_2O$ were conducted at different trap temperatures between 289 and 296 K, although the kinetic results obtained were always within the experimental error bars. Similarly, to check for possible differences brought about by the water vapour delivery method, several experiments were also performed at 296 K and at 127 K with $H_2O$ using the bubbler method. These measurements gave results that were essentially identical to those conducted with the CEM method.

The range of [$H_2O$] and [$D_2O$] that could be used in these experiments was limited by the formation of clusters in the supersonic flow, particularly at the lowest temperatures. Our earlier work on the kinetics of the C + $H_2O$/$D_2O$ reaction[20] allowed us to establish the appropriate ranges of [$H_2O$] and [$D_2O$] that could be used without significant interference from cluster formation. Moreover, earlier work on $H_2O$ cluster formation in He by Bourgalais et al.[44] suggest that cluster formation is likely to be negligible at all temperatures for the $H_2O$/$D_2O$ concentration ranges used during these studies.

$O(^1D)$ atoms were produced by the 10 Hz pulsed laser photolysis of ozone ($O_3$) at 266 nm with an average pulse energy of 22 mJ. At this wavelength, $O_3$ photodissociation results in high yields of $O(^1D)$ atoms.[15] Minor photodissociation products include $O(^3P)$, although these atoms are unreactive with $H_2O$ and $D_2O$ at low and ambient temperature. $O_3$ itself was created upstream of the Laval nozzle reservoir through the irradiation of a small flow of molecular oxygen in a quartz photolysis cell by a mercury lamp. In this process, $O(^3P)$ atoms produced by the UV photodissociation of $O_2$ reacted with $O_2$ to form $O_3$. The photolysis cell was maintained close to atmospheric pressure, to promote this termolecular association reaction. $O(^1D)$ was detected directly by VUV laser induced fluorescence (VUV LIF) in this study through its $2s^22p^4\ ^1D - 2s^22p^3(^2D°)3s\ ^1D°$ transition at 115.215 nm. For this purpose, a UV laser was focused into a cell containing a mixture of Xe and Ar, allowing tunable VUV radiation to be generated by frequency tripling. A more detailed description of this procedure can be found in earlier work.[2, 4, 8, 12, 25-28]

To reduce the detection of scattered light by the detector, the tripling cell was attached to the reactor through a long sidearm (75 cm). While a $MgF_2$ lens at the exit of the cell served to collimate the VUV beam, most of the residual divergent UV radiation was prevented from reaching the reactor by a series of baffles placed along the sidearm. Further attenuation of the VUV beam by residual gases within the sidearm was prevented by continuous flushing of this region by a secondary $N_2$ or Ar flow. The VUV probe beam was allowed to cross the supersonic flow at right angles, while on-resonance emission from unreacted $O(^1D)$ atoms within the flow was collected at right angles to both the flow itself and the probe beam by a solar blind photomultiplier tube (PMT). To avoid damage through exposure to reactive gases, the PMT was isolated from the reactor by a LiF window. The region between this window and the PMT contained a LiF lens to focus the VUV emission onto the PMT photocathode and was evacuated to prevent additional VUV absorption losses by atmospheric $O_2$. The PMT signal output was fed into a boxcar integration system that allowed the VUV LIF intensity to be measured as a function of delay time between the photolysis and probe lasers. 30 probe laser shots were averaged at any given time delay, with at least 70 time points typically recorded for each decay profile. As scattered light and other electronic noise could contribute to the recorded signal intensity, the baseline level was established by making several measurements at negative time delays.

Calibrated mass-flow controllers were used to regulate the gas flows during these experiments. Carrier gas Ar (99.999%), flush gas $N_2$ (99.999%), precursor gas $O_2$ (99.999%) and tripling gas Xe (99.998%) were flowed directly from cylinders. $D_2O$ with a purity of 99.8 % was employed, while a commercial ion-exchange resin system was used to obtain demineralized $H_2O$.

## 3 Theoretical Methods

### 3.1 Potential Energy Surface Calculations

The potential energy surface used in this work is based on the ground state potential of the $H_2O_2$ system.[45] Coelho and Brandão have enhanced the original PES in order to accurately describe the abstraction pathway of the $O(^1D) + H_2O$ reaction.[46] Despite being based on a large number of nuclear configurations, that potential was mainly focused on the equilibrium geometry and the different dissociation channels, mostly neglecting the abstraction pathway. By carrying out new *ab initio* calculations in 3741 nuclear configurations in this region, these authors have revised their $H_2O_2$ ($X\ ^1A$) PES.[46] In this new potential, the $O(^1D) + H_2O$ reaction can proceed by abstraction of the H-atom without an energy barrier. Despite that, in this enhanced PES there is not a different minimum energy path for the abstraction reaction and the minimum energy path for the $O(^1D) + H_2O$ reaction goes through the bottom of a well belonging to the $H_2O_2$ intermediate. Therefore, both reactions have a barrierless energy profile following the path: reactants ($O(^1D) + H_2O/D_2O$) → intermediate 1 ($H_2O$--$O$/$D_2O$--$O$, oxywater/oxy-heavy-water) → intermediate 2 ($H_2O_2/D_2O_2$, hydrogen peroxide/deuterium peroxide) leading to the formation of the products (2 OH/OD) (see Figure 4 of Ref. 44). Both intermediates 1 and 2 have lower potential energies than the reactants.

### 3.2 Quasiclassical (QCT) calculations

QCT trajectories have also been carried out for the $O(^1D) + H_2O$ and $O(^1D) + D_2O$ reactions based on this enhanced PES, at the temperatures of the experiment. Here, we used the program Venus96[47] adapted for this study.

In these calculations the $H_2O$ molecule is initialized at v=0 with its rotational quantum number and collision energy randomly generated from the corresponding Boltzmann distributions at the temperatures used in these calculations. For each temperature, the maximum impact parameter, $b_{max}$, has been defined as the impact distance where there was no reaction in a batch of 1000 trajectories.

### 3.3 Ring Polymer Molecular Dynamics (RPMD)

Based on an *ad hoc* extension of the imaginary-time path integral formalism, the RPMD method provides an efficient alternative to less reliable approaches based on conventional transition state theory (TST) and to more rigorous ones such as computationally expensive quantum mechanical (QM) dynamics methods for calculating thermal rate constants. Over the last decade, RPMD has been benchmarked extensively for numerous gas-phase chemical reactions.[48] In the RPMD method, the system is represented by a necklace of its *n* classical copies (beads) to map a quantum system.[49] These beads are connected to their nearest neighbors through a harmonic potential in an extended *n*-dimensional phase space.[50] The RPMD method provides a very reliable estimation of the correlation function for the thermal rate constants.[48] Several studies reported that the rate constant obtained within the RPMD formalism are exact within the high-temperature regime and only show a deviation of around a factor of 2-4 at low temperatures.[8, 26, 27, 31, 34, 48, 51-53] Since the RPMD method accurately treats the quantum Boltzmann operator, it can precisely map zero-point energy (ZPE) effects along the entire reaction pathway.[54] Moreover, it has better accuracy than other TST-based approximate methods in the deep quantum tunneling regime, as demonstrated in one of the stress-test scenarios.[55] Due to such impressive efficiency and simplicity, the RPMD method has been applied to many gas-phase bimolecular reactions as well as for condensed-phase processes. For example, see Refs. 8, 26, 27, 31, 34, 48, 51-53 and 56 and the references cited therein.

Mathematically, for a gas phase bimolecular reaction, the RPMD rate constant ($k_{RPMD}(T)$) at temperature $T$ can be written in terms of the Bennett-Chandler factorization scheme as:[57]

$$k_{RPMD}(T) = k_{QTST}(T; \xi^{\ddagger})\kappa(t \to \infty; \xi^{\ddagger}) \quad (2)$$

In Equation (2), $k_{QTST}(T; \xi^{\ddagger})$ is the centroid-density quantum transition-state theory (QTST) rate constant[58] evaluated at the reaction coordinate $\xi^{\ddagger}$. In this work, this factor has been calculated from the centroid potential of mean force (PMF) along the reaction coordinate.[57] $\kappa(t \to \infty)$ is the long-time limit of the time-dependent ring polymer transmission coefficient or ring polymer recrossing factor $\kappa(t)$. This factor is essentially a dynamical correction to the centroid density QTST rate constant, which guarantees that the final RPMD rate constant is independent of the choice of dividing surface. In the present work, this factor has been evaluated at the top of the free energy barrier so as to minimize the computation time required to attain a plateau value of $\kappa(t)$. The analytical PES used in the present RPMD study is derived from the modified PES of Coelho and Brandão[46] as described above. The RPMD calculations have been performed using the RPMDrate code[57] at the temperatures considered in the experimental part of the present study, and the other relevant input parameters are supplied in Table S1 of the electronic supplementary information (ESI).

## 4 Results

### 4.1 Experimental Results

All the experiments described here were performed with a large excess concentration of $H_2O$ or $D_2O$ with respect to $O(^1D)$, so that the pseudo-first-order approximation was valid. Under these conditions, the $O(^1D)$ atom concentration (which was considered to be proportional to the $O(^1D)$ VUV LIF signal) decreased exponentially as a function of time. Two such decay profiles recorded at 296 K for the $O(^1D) + H_2O$ reaction are shown in Figure 1.

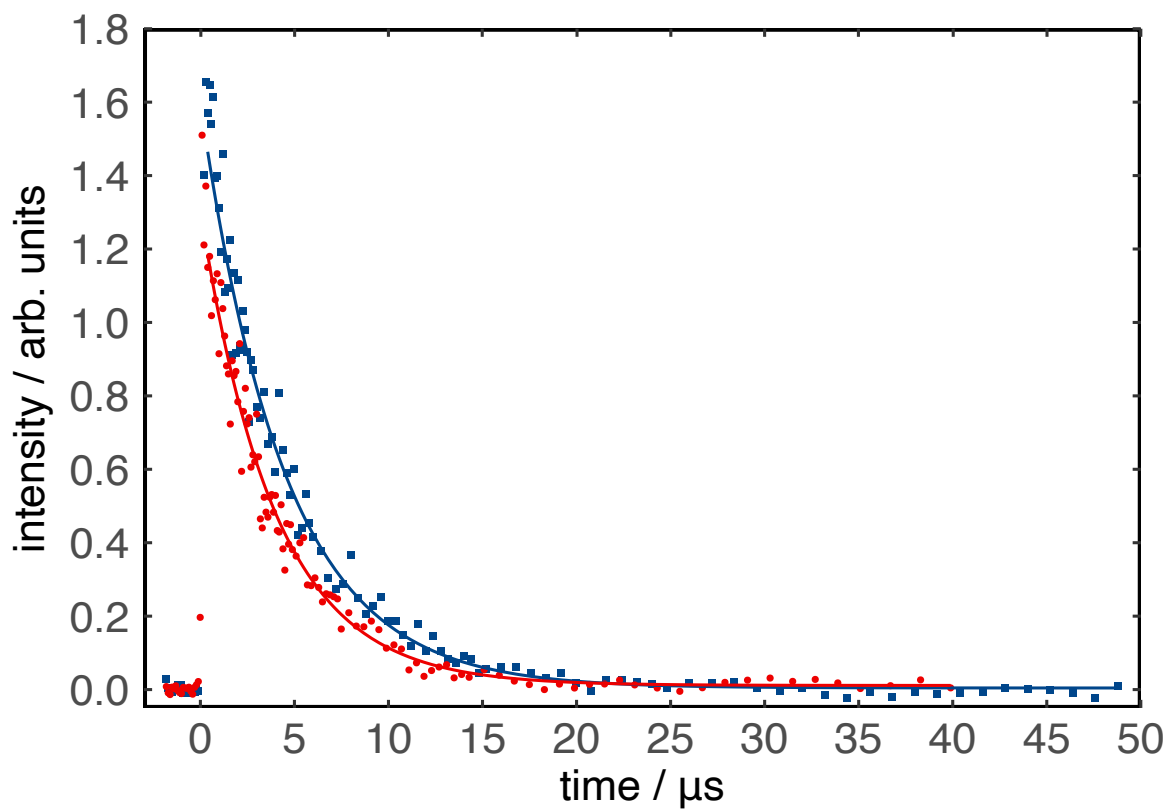

**Figure 1** $O(^1D)$ VUV LIF signal as a function of time between the photolysis and probe lasers, recorded at 50 K. (Blue solid squares) without $H_2O$ (the $O(^1D)$ VUV LIF signal decays due to quenching collisions of $O(^1D)$ with Ar); (red solid circles) $[H_2O] = 8.5 \times 10^{13}$ cm$^{-3}$. Solid lines represent single exponential fits to the individual datasets.

In the absence of $H_2O$ (or $D_2O$), $O(^1D)$ atoms are still removed rapidly from the flow due to electronic quenching with Ar. Nevertheless, the observed decay rate clearly increases upon the addition of water vapour to the flow. The intensity versus time profiles shown in Figure 1 can be well reproduced by the expression

$$I_{O(^1D)} = I_{O(^1D)_0} exp(-k_{1st}t) \qquad (3)$$

where $I_{O(^1D)_0}$ and $I_{O(^1D)}$ are the initial and time dependent VUV LIF intensities respectively, $k_{1st}$ is the pseudo-first-order rate constant and $t$ is the time. Only $O(^1D)$ losses by quenching and by reaction with $H_2O$ or $D_2O$ make significant contributions to the overall decay rate as previously discussed.[2] Solid lines in Figure 1 represent functional fits to the individual datasets, allowing values of $k_{1st}$ to be extracted. At least 18 individual decays similar to those presented in Figure 1 were recorded for a minimum of six different $H_2O$ or $D_2O$ concentrations at each temperature. The derived $k_{1st}$ values at a particular temperature were then plotted as a function of $[H_2O]$ or $[D_2O]$, as shown in Figure 2.

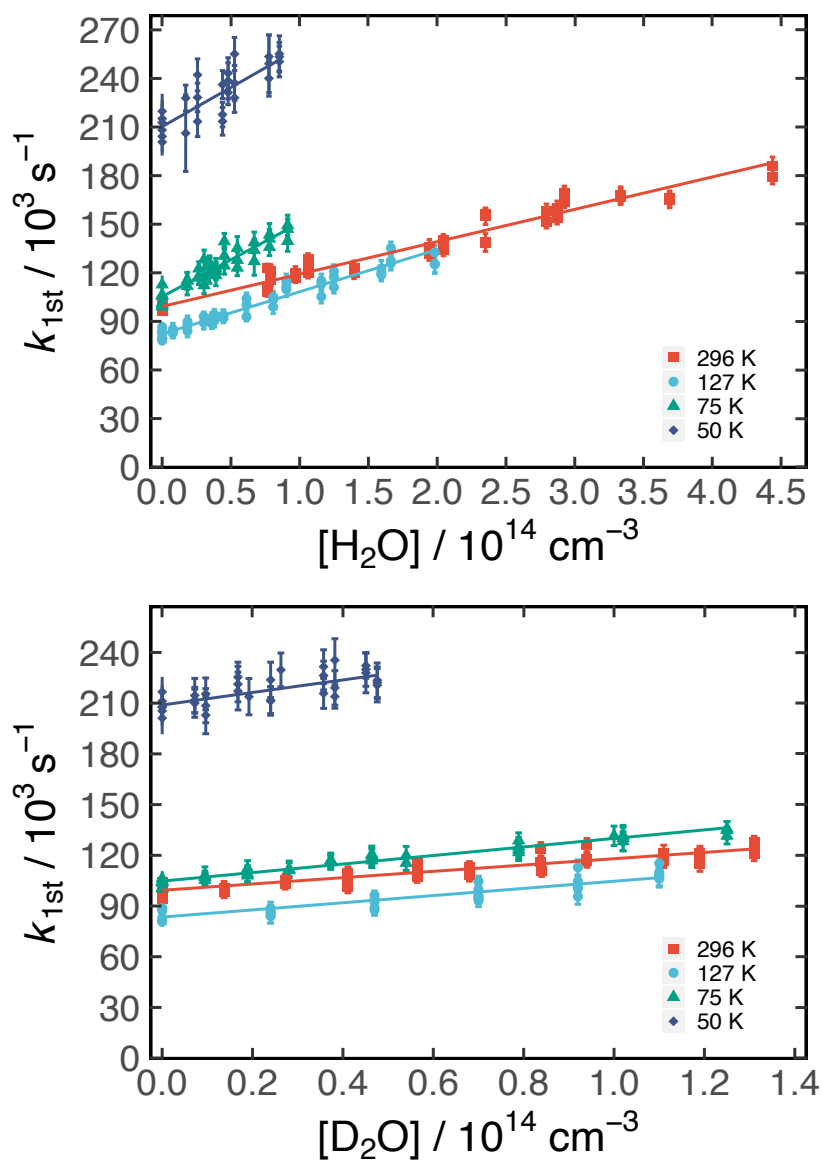

**Figure 2** Pseudo-first-order rate constant as a function of the $X_2O$ concentration, where X = H or D. (Upper panel) The $O(^1D)$ + $H_2O$ reaction. (Red solid squares) data recorded at 296 K; (blue solid circles) data recorded at 127 K; (green solid triangles) data recorded at 75 K; (blue solid diamonds) data recorded at 50 K. (Lower panel) The $O(^1D)$ + $D_2O$ reaction. The colours and symbols used are identical to those used in the upper panel. Solid lines represent weighted fits to the individual datasets with statistical uncertainties (1σ) derived from single exponential fits to intensity profiles similar to those shown in Figure 1.

Second-order rate constants were determined from the slopes of weighted linear least-squares fits to the data. The large y-axis intercept values of these plots arise from $O(^1D)$ quenching by the carrier gas. The much larger intercept value for the 50 K data is due to the increased Ar flow density at this temperature (2.59 × 10$^{17}$ cm$^{-3}$) compared with those data obtained at higher temperatures which are characterized by flow densities in the range (0.94-1.47) × 10$^{17}$ cm$^{-3}$. The weighting parameter used in this instance was the standard deviation (σ) obtained by the exponential fitting procedure described above, with each individual datapoint in Figure 2 weighted as 1/σ$^2$. The measured second-order rate constants for the O($^1$D) + H$_2$O and O($^1$D) + D$_2$O reactions are presented in Figure 3 alongside previous work. These values are summarized in Table 1 in addition to other relevant information.

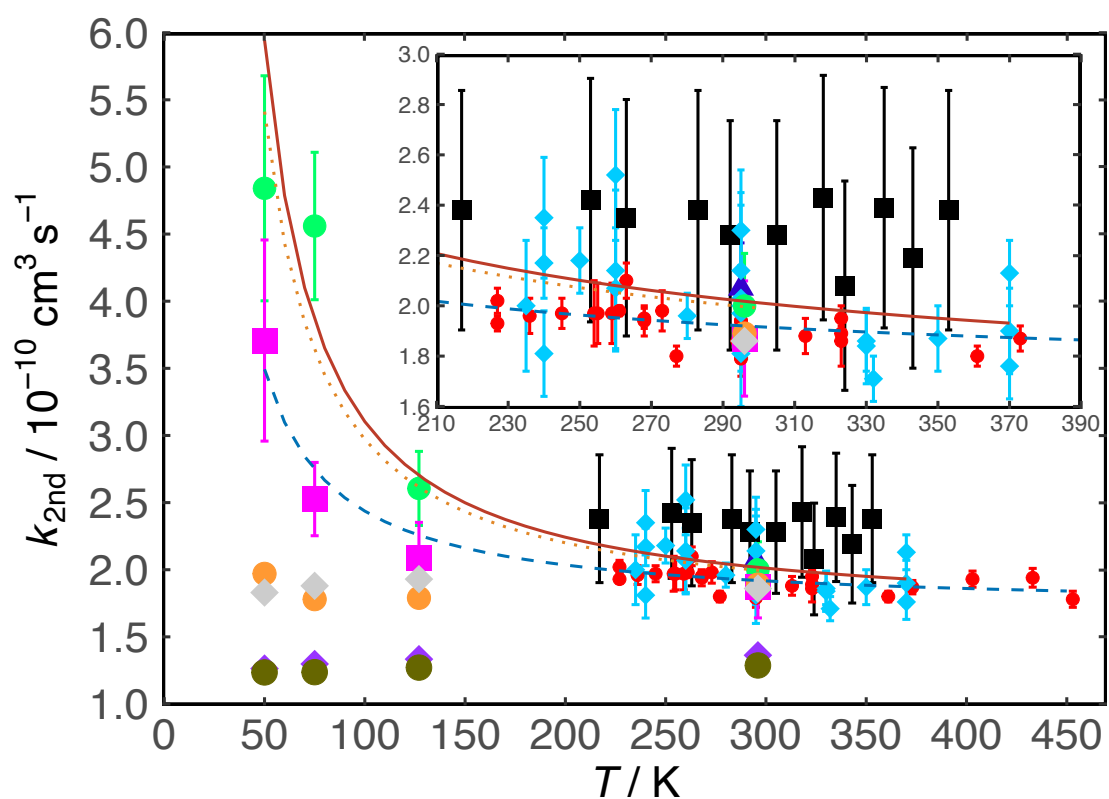

**Figure 3** Second-order rate constants for the O($^1$D) + H$_2$O and O($^1$D) + D$_2$O reactions as a function of temperature. The O($^1$D) + H$_2$O reaction. (Green circles) this experimental work; (black squares) Streit et al.[13]; (light blue diamonds) Dunlea et al.[14]; (red circles) Vranckx et al.[6]; (dark blue triangle) Takahashi et al.[59]; (orange circles) this work RPMD calculations; (purple diamonds) this work QCT calculations. (Solid brown line) Dunlea et al. Arrhenius fit[14]; (dashed blue line) Vranckx et al. Arrhenius fit[6]; (dotted brown line) Sander et al. recommendation.[15] The O($^1$D) + D$_2$O reaction. (Purple squares) this experimental work; (grey diamonds) this work RPMD calculations; (olive green circles) this work QCT calculations. Error bars on the present measurements represent the combined statistical (1σ) and estimated systematic uncertainty

(10 % of the nominal rate constant). The inset shows an expanded view of the congested region between 210 K and 390 K.

**Table 1** Measured second-order rate constants for the O($^1$D) + H$_2$O and O($^1$D) + D$_2$O reactions

| T / K | N[b] | [H$_2$O] / 10$^{13}$ cm$^{-3}$ | $k_{O(^1D)+H_2O}$ / 10$^{-10}$ cm$^3$ s$^{-1}$ | N[b] | [D$_2$O] / 10$^{13}$ cm$^{-3}$ | $k_{O(^1D)+D_2O}$ / 10$^{-10}$ cm$^3$ s$^{-1}$ | $k_H / k_D$ |
|---|---|---|---|---|---|---|---|
| 296 | 45 | 0 - 44.3 | (2.00 ± 0.21)[c] | 36 | 0 - 13.1 | (1.87 ± 0.23)[c] | 1.07 |
| 127 ± 2[a] | 50 | 0 - 19.8 | (2.61 ± 0.28) | 36 | 0 - 11.4 | (2.09 ± 0.27) | 1.25 |
| 75 ± 2 | 18 | 0 - 9.13 | (4.56 ± 0.55) | 34 | 0 - 12.5 | (2.53 ± 0.27) | 1.81 |
| 50 ± 1 | 26 | 0 - 8.5 | (4.84 ± 0.84) | 32 | 0 - 4.76 | (3.71 ± 0.75) | 1.30 |

[a]Uncertainties on the calculated temperatures represent the statistical (1σ) errors obtained from Pitot tube measurements of the impact pressure. [b]Number of individual measurements. [c]Uncertainties on the measured rate constants represent the combined statistical (1σ) and estimated systematic (10%) errors.

**4.2 QCT Results**

In Table 2, we present the main conditions used in the QCT calculations and the results obtained. Here, the electronic partition function ($Q_{el}$ = 5) has been taken into account when computing the rate constants, but not in the calculation of the reactive cross sections, $\sigma_r$. The quoted errors refer only to statistical errors in the computational procedure and do not consider the resonance and tunnelling effects that are ignored in the classical equations of motion.

**Table 2** Summary of the quasiclassical trajectory results obtained in this study

| | T / K | $b_{max}$ / Å | $N_{total}$[a] | $N_{react}$[b] | $\sigma_r$ / Å$^2$ | k / 10$^{-10}$ cm$^3$ s$^{-1}$ [c] | $k_H / k_D$ |
|---|---|---|---|---|---|---|---|
| O($^1$D)+H$_2$O | 50 | 14.0 | 39586 | 11331 | 176.3±1.4 | 1.25±0.01 | 1.03±0.01 |
| O($^1$D)+D$_2$O | 50 | 14.0 | 37925 | 10747 | 174.5±1.4 | 1.20±0.01 | |
| O($^1$D)+H$_2$O | 75 | 13.0 | 14612 | 4125 | 149.9±2.0 | 1.30±0.02 | 1.05±0.02 |

| | | | | | | |
|---|---|---|---|---|---|---|
| O($^1$D)+D$_2$O | 75 | 13.5 | 12523 | 3201 | 146.4±2.2 | 1.24±0.02 | |
| O($^1$D)+H$_2$O | 127 | 12.0 | 18775 | 4914 | 118.4±1.5 | 1.33±0.02 | 1.05±0.02 |
| O($^1$D)+D$_2$O | 127 | 12.0 | 17082 | 4363 | 115.6±1.5 | 1.27±0.02 | |
| O($^1$D)+H$_2$O | 296 | 10.5 | 30109 | 6891 | 79.3±0.8 | 1.36±0.01 | 1.06±0.02 |
| O($^1$D)+D$_2$O | 296 | 10.0 | 31626 | 7717 | 76.7±0.8 | 1.29±0.01 | |

[a]Total number of trajectories performed. [b]Number of reactive trajectories. [c]total rate constant divided by the electronic partition function.

The large impact parameter, $b_{max}$, is a result of the contribution of the intermolecular forces and the absence of an energy barrier for this reaction. As expected from reactions dominated by long-range forces, the reactive cross section decreases with temperature. Despite that, and in contrast to the experiment, we found a slight increase of the rate constant with temperature, which can be explained by the increase of the relative velocity of the reactants. The cross sections for the reactions involving H atoms in X$_2$O are somewhat larger than those for the corresponding reaction involving D atoms. In addition, the larger collision frequency for the H-atom reactions (lighter reduced mass) increases the effect of the cross sections and justifies the small isotopic effect. As the quasiclassical trajectories cannot reproduce the quantum tunnelling effects that might dominate in the abstraction process, and despite the combination of these two dynamical effects, the isotopic effect displayed in this table is smaller than the experimental findings.

To attain some insight on the reaction mechanism, we measured the time a trajectory spent at regions of the potential below -0.38 Eh (half the value between the products, -0.34 Eh, and the bottom of the intermediate H$_2$O$_2$ well, -0.42 Eh). The percentage of the reactive trajectories that do not reach this region of the potential or spend less than 50 fs is shown in Table 3.

**Table 3** Reactive trajectory behaviour at short times for the O($^1$D) + X$_2$O reactions

| | $T$ / K | 0 fs | < 50 fs |
|---|---|---|---|
| O($^1$D)+H$_2$O | 50 | 5.8 % | 67.6 % |
| O($^1$D)+D$_2$O | 50 | 5.0 % | 58.2 % |
| O($^1$D)+H$_2$O | 75 | 6.2 % | 67.0 % |
| O($^1$D)+D$_2$O | 75 | 5.2 % | 57.3 % |

| O($^1$D)+H$_2$O | 127 | 6.0 % | 66.1 % |
| O($^1$D)+D$_2$O | 127 | 5.2 % | 57.9 % |
| O($^1$D)+H$_2$O | 296 | 6.3 % | 67.6 % |
| O($^1$D)+D$_2$O | 296 | 4.7 % | 57.3 % |

We can conclude that most of the trajectories pass a very short time in such regions. This fraction seems to be independent of the temperature but there is a consistent difference between the behaviour of the reactive collisions with H$_2$O and those with D$_2$O, with the latter ones seeming to go deeper in the potential well.

### 4.3 RPMD results

The variation of PMF ($W(\xi)$) along the reaction coordinate $\xi$ for both O($^1$D) + H$_2$O and O($^1$D) + D$_2$O reactions are plotted in Figure S1 of the SM. From these plots, it is clear that both of these reactions show typical characteristics of a barrierless complex-forming reaction via intermediates 1 and 2. However, one can also detect small thermodynamic barriers just before entering the complex-forming zone, which increase with the rise in temperature. For example, in the case of the O($^1$D) + H$_2$O reaction, at 50 K, the barrier height is calculated to be 0.49 meV which increases to 15.52 meV at 296 K. Similarly, for the O($^1$D) + D$_2$O reaction, the barrier height increases from 1.44 meV (50 K) to 21.44 meV at 296 K. Apparently, this trend originates from a decrease in entropy from the reactant side to the intermediate complex-forming zone. Furthermore, the barrier height is always greater for the reaction involving D$_2$O. This fact is true for all the temperatures considered here and around the complex-forming zone and naturally becomes larger with increasing temperature (see inset plots of Fig. S1). For example, at 296 K, the PMF difference between O($^1$D) + H$_2$O and O($^1$D) + D$_2$O reactions in and around the complex-forming zone is 5.91 meV, while the same difference at 50 K is found to be 0.97 meV.

The time-dependent recrossing factors, $\kappa(t)$, for both reactions and all temperatures considered in this study are presented in Fig. S2, and the corresponding plateau values are reported in Table S2 of the ESI. It is evident that a considerable amount of recrossing takes place at all temperatures, which significantly alters the final rate constant value. A relatively long propagation time (0.5 ps) is required for the convergence of the $\kappa(t)$ value since both the reactions proceed through the formation of an intermediate complex.[48] One can observe initial oscillations in the $\kappa(t)$ values in both reactions existing at all temperatures considered in this study. These oscillations are the manifestation of the choice of the dividing surface and may correspond to O-O stretching vibrations on the PES[51, 52] that leads to the formation of the oxywater/oxy-heavy-water intermediates.[45] This vibration with a period of around ~38 fs (886 cm$^{-1}$)[60] leads to the repeated recrossing within the complex-forming zone. For both reactions, recrossing increases with temperature. For example, in the O($^1$D) + H$_2$O reaction, at 296 K, the plateau value of $\kappa(t)$ is 0.29, while at 50 K, it increases to 0.44. This fact may be due to the increased capability of the RPMD "child" trajectories to decay back to the reactant channel at higher temperatures and is also related to the barrier height in the PMF profile.[51, 53] Similar arguments can be put forward for the smaller recrossings observed at a particular temperature for the heavier deuterated reactants that have a more significant free energy barrier (see Table S2). The QTST and RPMD rate constant values are also reported in Table S2 of the ESI. Note that the RPMD rate constants have been corrected by including the electronic partition function contribution ($Q_{el}$ = 5).

## 5 Discussion

There are three earlier measurements of the rate constants for the O($^1$D) + H$_2$O reaction over a range of temperatures. Streit et al.[13] used the pulsed laser photodissociation of O$_3$ at 266 nm to produce O($^1$D) in their cryogenically cooled photolysis cell apparatus, following the time resolved decay of these atoms by emission via the weak spin-forbidden O($2^1$D$_2$) → O($2^3$P$_{3/2}$) transition at 630 nm. They obtained rate constants that were temperature independent over the 217-353 K range with a mean value of 2.3 × 10$^{-10}$ cm$^3$ s$^{-1}$. Dunlea et al.[14] used the pulsed laser photolysis of O$_3$ at 248 nm to produce O($^1$D) in their experiments, while O($^1$D) atoms were followed indirectly by following O($^3$P) atom production at 131 nm using a microwave discharge lamp and by fitting the resulting temporal profiles using a biexponential function. They derived rate constants over the 235-370 K range that increased slightly as the temperature fell. By combining their data with those of Streit et al.[13] they obtained the Arrhenius parameters A = (1.62 ± 0.27) × 10$^{-10}$ cm$^3$ s$^{-1}$ and E$_a$/R = (-65 ± 50) K. Vranckx et al.[6] used the pulsed laser photolysis of N$_2$O at 193 nm as the source of O($^1$D) in their experiments. O($^1$D) atoms were followed indirectly during this work using the O($^1$D) + C$_2$H → CH(A) + CO reaction and the detection of the CH(A → X) emission at 431 nm, effectively employing a chemical tracer method. In common with Dunlea et al.,[14] they measured rate constants with a very slight negative temperature dependence yielding the Arrhenius parameters A = (1.70 ± 0.12) × $10^{-10}$ $cm^3$ $s^{-1}$ and $E_a$/R = (-36 ± 20) K. The Arrhenius parameters recommended by the NASA panel evaluation on chemical kinetics and photochemical data for use in atmospheric studies by Sander et al.[15] are A = 1.63 × $10^{-10}$ $cm^3$ $s^{-1}$ and $E_a$/R = -60 K.

The room temperature rate constant value of $k_{O(^1D)+H_2O}(296\ K)$ = (2.00 ± 0.21) × $10^{-10}$ $cm^3$ $s^{-1}$ derived in the present work is in good agreement with the room temperature values derived by these three earlier studies. Other previous room temperature determinations[61-64] with a variety of monitoring techniques lie within the overall range (1.95-2.6) × $10^{-10}$ $cm^3$ $s^{-1}$ at room temperature. It should be noted though, that these previous studies followed the O($^1$D) kinetics through weak spin-forbidden transitions[13] or indirect detection methods[14, 61-64] in contrast to the direct method employed here. The only previous work to have used the same direct VUV LIF detection method was the study by Takahashi et al.,[59] who obtained a rate constant of (2.07 ± 0.18) × $10^{-10}$ $cm^3$ $s^{-1}$ at 295 K.

Above 100 K, the rate constant for the O($^1$D) + $H_2O$ reaction shows only a modest increase with decreasing temperature, reaching a value of $k_{O(^1D)+H_2O}(127\ K)$ = (2.61 ± 0.28) × $10^{-10}$ $cm^3$ $s^{-1}$. Below 100 K, the rate constant is seen to increase more rapidly reaching a value of $k_{O(^1D)+H_2O}(50\ K)$ = (4.84 ± 0.84) × $10^{-10}$ $cm^3$ $s^{-1}$. A similar trend is observed for the O($^1$D) + $D_2O$ reaction, albeit with a less pronounced increase below 100 K.

It can be seen from Figure 3 that the theoretical rate constants derived by the QCT method underestimate the measured values at all temperatures displaying very little variation as a function of temperature. These results are nonetheless in excellent agreement with the earlier room temperature QCT rate constant derived by Sayos et al.[18] of 1.32 × $10^{-10}$ $cm^3$ $s^{-1}$ based on an entirely different *ab initio* PES than the one used here. Although the experimental rate constants for the $H_2O$ reaction are significantly larger than those for the $D_2O$ reaction and diverge as the temperature falls, the QCT results for the H-atom variant remain 3-5 % larger than those derived for the D-atom variant over the entire temperature range. The greater kinetic isotopic effect found in the measured values is likely to be partly due to the fact that tunnelling is not accounted for in the H- / D- abstraction during quasiclassical trajectories.

Figure 3 indicates that the RPMD rate constants, $k_{RPMD}$, for both reactions do not vary much with temperature and are confined within interval of (1.78 – 1.97) × $10^{-10}$ $cm^3$ $s^{-1}$. Furthermore, RPMD calculations do not show any large kinetic isotope effect with the ratio $k_{RPMD}$(O($^1$D) + $H_2O$)/$k_{RPMD}$(O($^1$D) + $D_2O$) falling within the range 0.93 – 1.08. This may be due to the greater recrossing observed for the lighter hydrogen isotope that diminishes the sizeable kinetic isotope effect in the $k_{QTST}$ values (similar in magnitude with the experiment), resulting in no appreciable net kinetic isotope effect in the final $k_{RPMD}$ values. In terms of overall magnitude, the RPMD calculations reproduce well the present and previous experimental results at temperatures greater than 125 K, although below this temperature, the RPMD rate constants follow a similar temperature independent behaviour to those derived at higher temperatures (and the QCT results), in contrast to the large increase displayed by the experimental rates below 100 K. The maximum difference between the RPMD calculations and the experiment is a factor of 3, obtained at 50 K. However, this degree of accuracy is acceptable considering the accuracy level of the given PES and the RPMD formalism, particularly at low temperatures, as discussed previously in Section 3.3. At 296 K, the accuracy of the RPMD simulations reaches up to 75%. Furthermore, the RPMD calculation could also qualitatively capture the increase in the O($^1$D) + $H_2O$ reaction rate for the decrease in temperature from 75 K to 50 K. The $k_{RPMD}$ values seem to have a better agreement with the experimental observations than the results obtained by the QCT calculations within the temperature regime considered in this study.

The failure of the QCT and RPMD calculations to capture the rapid increase displayed by the experimental rate constants at low temperatures could have several origins. One possibility is that the non-adiabatic quenching process (1b), O($^1$D) + $H_2O$ → O($^3$P) + $H_2O$, becomes significantly more important as the temperature falls. The experiments reported here follow the total loss of O($^1$D) atoms as a function of time, which could potentially include a component due to quenching loss in addition to the reactive part. In contrast, both QCT and RPMD calculations only consider the reactive contribution occurring over the ground state $^1$A surface of the $H_2O_2$ system. Experiments could eventually be devised to quantify this quenching contribution, by following the OH product of the O($^1$D) + H$_2$O reaction and comparing with the OH formed by a reference process such as the O($^1$D) + CH$_4$ reaction.[12] Nevertheless, the OH product from both reactions is likely to be formed over a range of vibrational and/or rotation levels which would all need to be probed for a quantitative assessment of this effect, making such measurements difficult and time consuming. Another possibility would be to follow the increase of the O($^3$P) concentration as a function of time in a similar manner to the earlier experiments of Takahashi et al.,[16] who followed the quenching of O($^1$D) to O($^3$P) through collisions with both H$_2$O and D$_2$O at room temperature. These authors measured very small quenching contributions, $\Phi_q$ = (0.02 $\pm$ 0.01) for these two species when compared to the quenching yield with N$_2$ ($\Phi_q$ = 1). Although the quenching contribution is small around 300 K, this may not be the case at low temperatures and merits further investigation. Nevertheless, the VUV transitions typically used to follow O($^3$P) atoms around 130.2 nm are not accessible with the tripling method currently employed in our laboratory.

Another possible explanation for the observed divergence between experimental and theoretical results at low temperature could be due to the involvement of other potential surfaces. The present calculations consider that only a small fraction of the collisions occur over the ground $^1$A surface of the H$_2$O$_2$ system, leading to a 1/5 corrective factor in the calculation of the final rate constants. Indeed, if non-adiabatic transitions between some or all of the five PESs correlating with reagents O($^1$D) and H$_2$O occur at long range, and these surfaces are attractive, the actual factor could be significantly larger. In particular, it is expected that the calculated reaction rate constants would increase if the contribution from the first excited state ($^1$A″) is also considered.[18] Therefore, it is necessary to perform dynamical simulations on this PES, which is ignored in this present work due to the lack of availability of such PESs in the literature, to obtain a much better agreement with the experiment.

## 6 Conclusions

This paper reports the results of a joint experimental and theoretical study of the O($^1$D) + H$_2$O and O($^1$D) + D$_2$O reactions. Experimentally, the kinetics of these reactions were investigated between 50 K and 296 K using a supersonic flow reactor. O($^1$D) atoms were created by the pulsed laser photolysis of O$_3$ at 266 nm and detected by pulsed laser induced fluorescence at

115.2 nm. The rate constants for both reactions were measured to be large ( > 1.5 × 10$^{-10}$ cm$^3$ s$^{-1}$) at all temperatures, remaining relatively constant at temperatures above 100 K. Theoretically, new *ab initio* calculations were performed of the X $^1$A potential energy surface of the H$_2$O$_2$ system involved in the reaction, allowing us to provide a better description of the barrierless H-atom abstraction pathway when combined with earlier calculations. This enhanced surface was subsequently employed in quasiclassical trajectory calculations and ring polymer molecular dynamics simulations to obtain rate constants for both reactions. The calculated rate constants display a similar temperature independent behaviour to the experimental ones above 100 K, with the theoretical values derived by the ring polymer molecular dynamics method being closer to the measured values. In contrast, while the theoretical rate constants remain relatively constant below 100 K, the experimental values for both reactions increase significantly. Plausible explanations for the discrepancy between experiment and theory are presented, including a possible increased contribution of quenching losses during the low temperature measurements and/or the contribution of excited potential energy surfaces to the reaction which are not accounted for by the present calculations. The need for further experimental and theoretical work to elucidate the origin of this effect is highlighted.

**Conflicts of interest**

There are no conflicts to declare.

**Acknowledgements**

K. M. H. acknowledges support from the French program ''Physique et Chimie du Milieu Interstellaire'' (PCMI) of the CNRS/INSU with the INC/INP co-funded by the CEA and CNES as well as funding from the ''Programme National de Planétologie'' (PNP) of the CNRS/INSU.
This research was also supported by the EMME-CARE project that has received funding from the European Union's Horizon 2020 Research and Innovation Program, under Grant Agreement No. 856612, as well as matching co-funding by the Government of the Republic of Cyprus. Y. V. S. was also supported by the Russian Foundation for Basic Research grant number 20-03-00833. S. B. acknowledges the financial support of the European Regional Development Fund and the Republic of Cyprus through the Research Promotion Foundation

NANO²LAB Project INFRASTRUCTURES/1216/0070. The RPMD computations have been performed on the "AMD EPYC" High Performance Computing Facility of The Cyprus Institute. J.B. and D.V.C. acknowledge the support by the FCT under the PTDC/QUI-QFI/31955/2017 research project, cofinanced by the European Community Fund, FEDER and the COST Action CA18212 Molecular Dynamics in the GAS phase.

**References**

1. J. R. Wiesenfeld, *Acc. Chem. Res.*, 1982, **15**, 110-116.
2. R. Grondin, J.-C. Loison and K. M. Hickson, *J. Phys. Chem. A*, 2016, **120**, 4838-4844.
3. J. A. Schmidt, M. S. Johnson and R. Schinke, *PNAS*, 2013, **110**, 17691-17696.
4. D. Nuñez-Reyes and K. M. Hickson, *J. Phys. Chem. A*, 2018, **122**, 4002-4008.
5. C. S. Boxe, J. S. Francisco, R. L. Shia, Y. L. Yung, H. Nair, M. C. Liang and A. Saiz-Lopez, *Icarus*, 2014, **242**, 97-104.
6. S. Vranckx, J. Peeters and S. Carl, *Phys. Chem. Chem. Phys.*, 2010, **12**, 9213-9221.
7. J. A. Davidson, H. I. Schiff, G. E. Streit, J. R. McAfee, A. L. Schmeltekopf and C. J. Howard, *J. Chem. Phys.*, 1977, **67**, 5021-5025.
8. K. M. Hickson and Y. V. Suleimanov, *J. Phys. Chem. A*, 2017, **121**, 1916-1923.
9. S. Vranckx, J. Peeters and S. Carl, *Phys. Chem. Chem. Phys.*, 2008, **10**, 5714-5722.
10. M. A. Blitz, T. J. Dillon, D. E. Heard, M. J. Pilling and I. D. Trought, *Phys. Chem. Chem. Phys.*, 2004, **6**, 2162-2171.
11. T. J. Dillon, A. Horowitz and J. N. Crowley, *Chem. Phys. Lett.*, 2007, **443**, 12-16.
12. Q. Y. Meng, K. M. Hickson, K. J. Shao, J. C. Loison and D. H. Zhang, *Phys. Chem. Chem. Phys.*, 2016, **18**, 29286-29292.
13. G. E. Streit, *J. Chem. Phys.*, 1976, **65**, 4761-4764.
14. E. J. Dunlea and A. R. Ravishankara, *Phys. Chem. Chem. Phys.*, 2004, **6**, 3333-3340.
15. S. P. Sander, J. Abbatt, J. R. Barker, J. B. Burkholder, R. R. Friedl, D. M. Golden, R. E. Huie, C. E. Kolb, M. J. Kurylo, G. K. Moortgat, V. L. Orkin and P. H. Wine, *JPL Publication 10-6, Jet Propulsion Laboratory, Pasadena, 2011 http://jpldataeval.jpl.nasa.gov.*, 2011.
16. K. Takahashi, R. Wada, Y. Matsumi and M. Kawasaki, *J. Phys. Chem.*, 1996, **100**, 10145-10149.
17. R. Zellner, G. Wagner and B. Himme, *J. Phys. Chem.*, 1980, **84**, 3196-3198.
18. R. Sayós, C. Oliva and M. González, *J. Chem. Phys.*, 2000, **113**, 6736-6747.
19. K. M. Hickson, P. Caubet and J.-C. Loison, *J. Phys. Chem. Lett.*, 2013, **4**, 2843-2846.
20. K. M. Hickson, J.-C. Loison, D. Nuñez-Reyes and R. Méreau, *J. Phys. Chem. Lett.*, 2016, **7**, 3641-3646.
21. K. M. Hickson, *J. Phys. Chem. A*, 2019, **123**, 5206-5213.
22. N. Daugey, P. Caubet, B. Retail, M. Costes, A. Bergeat and G. Dorthe, *Phys. Chem. Chem. Phys.*, 2005, **7**, 2921-2927.
23. N. Daugey, P. Caubet, A. Bergeat, M. Costes and K. M. Hickson, *Phys. Chem. Chem. Phys.*, 2008, **10**, 729-737.
24. K. M. Hickson, P. Larrégaray, L. Bonnet and T. González-Lezana, *Int. Rev. Phys. Chem.*, in press.
25. D. Nuñez-Reyes and K. M. Hickson, *J. Phys. Chem. A*, 2018, **122**, 4696-4703.

26. D. Nuñez-Reyes, K. M. Hickson, P. Larrégaray, L. Bonnet, T. González-Lezana, S. Bhowmick and Y. V. Suleimanov, *J. Phys. Chem. A*, 2019, **123**, 8089-8098.
27. D. Nuñez-Reyes, K. M. Hickson, P. Larrégaray, L. Bonnet, T. González-Lezana and Y. V. Suleimanov, *Phys. Chem. Chem. Phys.*, 2018, **20**, 4404-4414.
28. D. Nuñez-Reyes, J. Kłos, M. H. Alexander, P. J. Dagdigian and K. M. Hickson, *J. Chem. Phys.*, 2018, **148**, 124311.
29. J. Bourgalais, M. Capron, R. K. A. Kailasanathan, D. Osborn, L. , K. Hickson, M. , J.-C. Loison, V. Wakelam, F. Goulay and S. Le Picard, D. , *Astrophys. J.*, 2015, **812**, 106.
30. K. M. Hickson, J.-C. Loison, J. Bourgalais, M. Capron, S. D. Le Picard, F. Goulay and V. Wakelam, *Astrophys. J.*, 2015, **812**, 107.
31. K. M. Hickson, J.-C. Loison, H. Guo and Y. V. Suleimanov, *J. Phys. Chem. Lett.*, 2015, **6**, 4194-4199.
32. K. M. Hickson, J.-C. Loison, F. Lique and J. Kłos, *J. Phys. Chem. A*, 2016, **120**, 2504-2513.
33. K. M. Hickson, J.-C. Loison and V. Wakelam, *Chem. Phys. Lett.*, 2016, **659**, 70-75.
34. K. M. Hickson and Y. V. Suleimanov, *Phys. Chem. Chem. Phys.*, 2017, **19**, 480-486.
35. D. Nuñez-Reyes and K. M. Hickson, *J. Phys. Chem. A*, 2017, **121**, 3851-3857.
36. D. Nuñez-Reyes and K. M. Hickson, *Chem. Phys. Lett.*, 2017, **687**, 330-335.
37. D. Nuñez-Reyes and K. M. Hickson, *Phys. Chem. Chem. Phys.*, 2018, **20**, 17442-17447.
38. D. Nuñez-Reyes, J.-C. Loison, K. M. Hickson and M. Dobrijevic, *Phys. Chem. Chem. Phys.*, 2019, **21**, 22230-22237.
39. D. Nuñez-Reyes, J.-C. Loison, K. M. Hickson and M. Dobrijevic, *Phys. Chem. Chem. Phys.*, 2019, **21**, 6574-6581.
40. R. J. Shannon, C. Cossou, J.-C. Loison, P. Caubet, N. Balucani, P. W. Seakins, V. Wakelam and K. M. Hickson, *RSC Adv.*, 2014, **4**, 26342-26353.
41. Y. Wu, J. Cao, H. Ma, C. Zhang, W. Bian, D. Nunez-Reyes and K. M. Hickson, *Sci. Adv.*, 2019, **5**, eaaw0446.
42. J. A. Davidson, H. I. Schiff, T. J. Brown, G. E. Streit and C. J. Howard, *J. Chem. Phys.*, 1978, **69**, 1213-1215.
43. N. Matsunaga and A. Nagashima, *Int. J. Thermophys.*, 1987, **8**, 681-694.
44. J. Bourgalais, V. Roussel, M. Capron, A. Benidar, A. W. Jasper, S. J. Klippenstein, L. Biennier and S. D. Le Picard, *Phys. Rev. Lett.*, 2016, **116**, 113401.
45. D. V. Coelho and J. Brandão, *Phys. Chem. Chem. Phys.*, 2017, **19**, 1378-1388.
46. D. V. Coelho and J. Brandão, unpublished work.
47. W. L. Hase, R. J. Duchovic, X. Hu, A. Komornicki, K. F. Lim, D. Lu, G. Peslherbe, K. Swamy, S. R. V. Linde, A. Varandas, H. Wang and R. J. Wolf, *Quantum Chemical Program Exchange Bulletin*, 1996, **16**, 43.
48. Y. V. Suleimanov, F. J. Aoiz and H. Guo, *J. Phys. Chem. A*, 2016, **120**, 8488-8502.
49. I. R. Craig and D. E. Manolopoulos, *J. Chem. Phys.*, 2004, **121**, 3368-3373.
50. D. Chandler and P. G. Wolynes, *J. Chem. Phys.*, 1981, **74**, 4078-4095.
51. S. Bhowmick, D. Bossion, Y. Scribano and Y. V. Suleimanov, *Phys. Chem. Chem. Phys.*, 2018, **20**, 26752-26763.
52. J. Espinosa-Garcia, M. Garcia-Chamorro, J. C. Corchado, S. Bhowmick and Y. V. Suleimanov, *Phys. Chem. Chem. Phys.*, 2020, **22**, 13790-13801.
53. T. González-Lezana, D. Bossion, Y. Scribano, S. Bhowmick and Y. V. Suleimanov, *J. Phys. Chem. A*, 2019, **123**, 10480-10489.
54. R. P. de Tudela, F. J. Aoiz, Y. V. Suleimanov and D. E. Manolopoulos, *J. Phys. Chem. Lett.*, 2012, **3**, 493-497.

55. R. Perez de Tudela, Y. V. Suleimanov, J. O. Richardson, V. Saez Rabanos, W. H. Green and F. J. Aoiz, *J Phys Chem Lett*, 2014, **5**, 4219-4224.
56. S. Bhowmick, M. I. Hernández, J. Campos-Martínez and Y. V. Suleimanov, *Phys. Chem. Chem. Phys.*, 2021, **23**, 18547-18557.
57. Y. V. Suleimanov, J. W. Allen and W. H. Green, *Comput. Phys. Commun.*, 2013, **184**, 833-840.
58. I. R. Craig and D. E. Manolopoulos, *J. Chem. Phys.*, 2005, **123**, 034102.
59. K. Takahashi, Y. Takeuchi and Y. Matsumi, *Chem. Phys. Lett.*, 2005, **410**, 196-200.
60. R. Sayós, C. Oliva and M. González, *J. Chem. Phys.*, 2001, **115**, 8828-8837.
61. S. T. Amimoto, A. P. Force, R. G. Gulotty and J. R. Wiesenfeld, *J. Chem. Phys.*, 1979, **71**, 3640-3647.
62. P. H. Wine and A. R. Ravishankara, *Chem. Phys. Lett.*, 1981, **77**, 103-109.
63. K. H. Gericke and F. J. Comes, *Chem. Phys. Lett.*, 1981, **81**, 218-222.
64. L. C. Lee and T. G. Slanger, *Geophys. Res. Lett.*, 1979, **6**, 165-166.